\documentclass[prl,amsmath,amssymb,superscriptaddress,twocolumn]{revtex4-1}
\usepackage{graphicx}
\usepackage[bookmarks=false]{hyperref}
\usepackage{multirow}

\DeclareMathOperator{\erf}{erf} % used in Equation 2
\DeclareMathOperator{\M}{\mathit M} % used in Equation 2

\begin{document}

\title{Reducing Disorder in Artificial Kagome Ice}

\author{Stephen A. Daunheimer}
\affiliation{Department\;of\;Materials\;Science\;\&\;Engineering,\;University\;of\;Maryland,\;College\;Park,\;Maryland\;20742,\;USA}
\author{Olga Petrova}
\author{Oleg Tchernyshyov}
\affiliation{Department of Physics \& Astronomy, Johns Hopkins
  University, Baltimore, Maryland 21218, USA}
\author{John Cumings}
\email{Corresponding author, cumings@umd.edu}
\affiliation{Department\;of\;Materials\;Science\;\&\;Engineering,\;University\;of\;Maryland,\;College\;Park,\;Maryland\;20742,\;USA}

\date{\today}

\begin{abstract}
  Artificial spin ice has become a valuable tool for understanding
  magnetic interactions on a microscopic level.  The strength in the
  approach lies in the ability of a synthetic array of nanoscale
  magnets to mimic crystalline materials, composed of atomic magnetic
  moments.  Unfortunately, these nanoscale magnets, patterned from
  metal alloys, can show substantial variation in relevant quantities
  such as coercive field, with deviations up to 16\%.  By carefully
  studying the reversal process of artificial kagome ice, we can
  directly measure the distribution of coercivities, and by switching
  from disconnected islands to a connected structure, we find that the
  coercivity distribution can achieve a deviation of only 3.3\%.
  These narrow deviations should allow the observation of behavior
  that mimics canonical spin-ice materials more closely.

\end{abstract}

\maketitle

Water ice and spin ice are classic examples of geometrically
frustrated systems \cite{diep, IFM}, both with residual
low\nobreakdash-$T$ entropy \cite{nature.399.333}.  In water ice,
thermodynamic phases with ordered protons were discovered after
decades of experiments \cite{petrenko}.  In contrast, no
dipole-ordered phase has been observed in spin ice even at the lowest
accessible temperatures, contrary to a theoretical prediction
\cite{melko2001}.  Divergent relaxation times and quenched disorder in
samples have been cited as possible explanations.  Artificial spin ice
has been proposed to help address these questions \cite{wang2006}, as
it allows the direct control of the geometry of the lattice, with the
combined ability to directly image the resulting microstate.  Here,
samples are composed of lattices of nanoscale ferromagnetic islands,
where the magnetization of each element points along its longitudinal
axis.  At the vertices of the lattice, the ferromagnetic elements
interact, and because of the geometry of the system, their magnetic
configurations are frustrated
\cite{wang2006,morgan2011,ladak2010,mengotti2011,tanaka2006,qi2008}.
This allows the study of frustration in systems where crystalline
imperfections can be completely removed by design, or introduced in a
controlled way.  Unfortunately, current lithographic techniques are
limited by unintended roughness at edges and interfaces, creating
inadvertent disorder.  This diminishes the ability to compare
observations from artificial spin ice materials with studies of spin
ice oxides, where magnetic atoms are presumed to be identical.

Edge roughness of nanomagnetic elements is known to substantially
influence the coercive field, by creating nucleation sites that can
initiate the magnetic reversal \cite{gadbois1995}.  In some artificial
spin ice geometries, this edge roughness can create a large
variability in the behavior of the artificial ``atoms'' (magnetic
nano-islands).  In recent studies of artificial kagome ice, the
variations in coercivity were found to be substantial---up to 16\% of
the average coercive value \cite{mengotti2011,ladak2010,kohli2011}.
This variability can easily be reduced by choosing materials with low
crystal anisotropy \cite{mengotti2011}, but we here show substantial
further reduction with a geometry with connected magnetic islands.  In
a connected geometry, nontrivial spin textures (domain walls) already
exist at the vertices even in equilibrium, and thus the reversal
process is not dictated by the nucleation of new domain walls, a step
that is more susceptible to the effects of edge roughness.

To address the problem of disorder and variability in artificial ice,
techniques must first be established to measure the static disorder in
the material, a problem that is complicated by the sought-after
occurrence of statistical disorder, in analogy with thermal disorder
in pyrochlore spin ice.  Prior work has addressed this problem by
focusing on the reversal of artificial kagome ice in an external field
applied at 180$^\circ$ to the direction of initial magnetization
\cite{mengotti2011,ladak2010}, a potentially deterministic process, in
which statistical variations are hopefully minimized.  However, the
reversal process under these conditions shows large discrete avalanche
events, and for the materials with small static disorder described
below, the variations in these reversals are almost entirely dominated
by non-reproducible statistical disorder.  Prior studies provided
disorder estimates as large as 16\% through Monte Carlo modeling of
the 180$^\circ$ reversal process, but this approach is not possible
when the amount of quenched disorder is much smaller.

In this Letter, we present a method to alleviate this issue by
performing reversals with the field applied at $120^{\circ}$ and
$100^{\circ}$ to the direction of initial magnetization instead of
$180^{\circ}$\!\!\!.\, Under these conditions, magnetization reversal
in individual islands proceeds largely independently of their
neighbors, preventing the formation of avalanches. This enables us to
extract the spread of coercivities directly from the magnetization
curves $\M(H)$.  These data are then used to calculate a disorder
parameter, $\sigma/\overline{{H}_{c}}$, where $\overline{{H}_{c}}$ is
the average coercive field of a magnetic element and $\sigma$ is the
standard deviation of the distribution.  Using this parameter, we are
able to directly compare our data with those from other groups.  By
considering carefully the magnetic reversal process, we demonstrate
that this disorder parameter can be greatly reduced in a system with
connected magnetic islands.  Additionally, all data used to determine
the disorder present in a crystal can be obtained directly from
experiments rather than relying on parameters extracted from Monte
Carlo simulations.

\begin{figure}
  \begin{center}
    \includegraphics[width=\columnwidth]{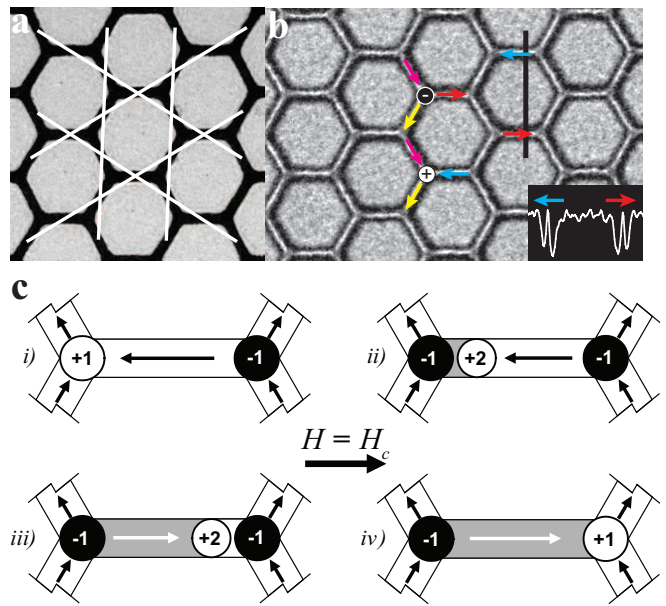}
    \caption{\label{fig1}(a) In-focus TEM image of the artificial
      kagome ice. Lines representing the kagome planes are overlaid to
      show how the origin of the honeycomb structure of our
      crystals. (b) Lorentz contrast image of the kagome lattice.  The
      contrast difference can clearly be seen with a bright and dark
      edge across each element in the array. The magnetization of six
      elements is shown along with the resulting charge at the vertex
      they are connected to. Inset: the intensity profile along the
      indicated black line. The magnetization of two elements is
      captured in the profile and depicted by arrows on the Lorentz
      image and on the profile itself. (c) A cartoon of the reversal
      process of an individual element is shown in panels
      (\textit{i})-(\textit{iv}), as described in the text.}
  \end{center}
\end{figure}

In our studies, we choose the kagome lattice for our artificial spin
ice \cite{ladak2010,mengotti2011,tanaka2006,qi2008}, as shown in
Fig.~\ref{fig1}(a) and \ref{fig1}(b).  Samples were patterned via
e-beam lithography onto an electron transparent SiN membrane with PMMA
resist \cite{qi2008}.  The Ni$_{80}$Fe$_{20}$ films were deposited by
e-beam evaporation followed by metal lift-off.  The resulting elements
are 500 nm long, 110 nm wide, and 23 nm thick with an additional 4 nm
capping layer of Al to prevent oxidation.  Reversals were performed
\textit{in situ} in a transmission electron microscope in Lorentz mode
to capture the magnetic configuration of the crystal throughout a
reversal.  A magnified image of a portion of the sample can be seen in
Fig.~\ref{fig1}(a).  Magnetic reversal in this system can be described
in terms of the emission, propagation, and absorption of domain walls
carrying non-zero magnetic charge \cite{mellado2010}.  Similar
emergent magnetic monopole excitations have been shown to exist in
both conventional
\cite{castelnovo2008,ryzhkin2005,jaubert2009,morris2009,fennell2009,bramwell2009}
and artificial spin ice \cite{ladak2010,mengotti2011}.

The nature of magnetic charges is understood as follows. Because
magnetic induction $\mathbf B = \mu_0(\mathbf H + \mathbf M)$ is
divergence-free, a region where lines of magnetization $\mathbf M$
terminate or originate becomes a source of sink of magnetic field
$\mathbf H$.  Thus magnetic charges in artificial spin ice contain
integer multiples of the magnetization flux of a single magnetic
element.  In these units, a monopole propagating along an element
during reversal can carry a charge $\pm2$, and a vertex can carry a
charge of $\pm1$ or $\pm3$.  The interactions between the charges via
$\mathbf H$ can be described simply in terms of a Coulomb potential
\cite{mellado2010}.  Triple charges represent ice-rule violations for
the kagome lattice, and although they have been observed
experimentally during 180$^\circ$ reversals in other systems
\cite{ladak2010,mengotti2011}, they \textit{never} appear in the
low-disorder system we study here \cite{qi2008}.

Before considering the angle dependence, it is instructive to consider
the microscopic details of the reversal process, depicted in Figure
\ref{fig1}(c).  In panel (\textit{i}), the horizontal element will be
the first to reverse because the external magnetic field has the
largest component along its axis. The reversal begins with the
emission of a domain wall of $+2$ charge into the horizontal element,
leaving behind a vertex with $-1$ charge.  The domain wall moves along
the element until it reaches the opposite vertex, thereby reversing
the element's magnetization.  In panel (\textit{iii}), the vertex with
charge $-1$ pulls the domain wall with charge $+2$ in, thereby
changing its own charge to $+1$.  The reversal is different for the
disconnected lattice where a domain wall needs to be nucleated at the
end of an element, rather than emitted from a vertex.  In this case,
the field required to inject a charged domain wall can vary
significantly depending on crystal symmetry and edge roughness of the
elements.
\begin{figure}
  \begin{center}
    \includegraphics[width=\columnwidth]{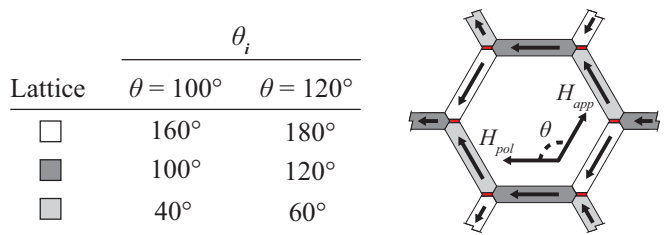}
    \caption{\label{fig2}Illustration of the crystal magnetization at
      the beginning of a $\theta=120^{\circ}$ reversal. The table
      shows values of $\theta_{i}$ for the elements of the three
      sublattices. The angles are with respect to $H_{app}$. The red
      lines at the vertices indicate the location of a head-to-head or
      tail-to-tail domain wall which is the source of magnetic charge
      at each vertex.}
   \end{center}
\end{figure}

One might expect the magnitude of the coercive field to be lowest when
the field is parallel to the element receiving the emitted domain
wall: only the longitudinal component of the magnetic field pushes a
domain wall along the element.  However, in a connected lattice, the
transverse field component also plays a role, thanks to an asymmetric
distribution of magnetic charge within each vertex. 

Consider a simplified model of a vertex consisting of three thin
domain walls meeting at the center of the vertex, Fig.~\ref{fig2}. In
a vertex with unit charge, two of the walls are neutral (head-to-tail)
and one is magnetically charged (head-to-head or tail-to-tail). An
applied field exerts a maximum force on the charged domain wall when
it is applied normal to the wall, or at $30^\circ$ to the axis of the
element receiving the wall. Micromagnetic simulations confirm these
qualitative considerations \cite{mellado2010}.  The coercive field
depends on its orientation as follows:
\begin{equation} \label{right}
    H_{c}(\theta_i)=H^{i}_{c}/\lvert\cos\left(\theta_i+\alpha\right)\rvert.
\end{equation}
Here $\theta_i$ is the angle of the applied field with the long axis
of the $i$\textsuperscript{th} element, $H^{i}_{c}$ is the minimum
coercivity of that element, and $\alpha = 19^\circ$ is an offset angle
\cite{mellado2010}.

Using this model, we measure the disorder in our artificial ice, as
caused by variability in the coercive field of the elements, by
performing magnetic reversals of large arrays with more than $10^4$
elements. The system is initially magnetized in a high field $H \gg
\overline{{H}_{c}}$. The field is switched off and the sample is
rotated through the desired angle $\theta$ in its plane. In this
Letter, we focus on $\theta = 100^\circ$ and $120^\circ$\!\!\!.\, As
can be seen from Eq.~\eqref{right}, these orientations guarantee large
differences in the coercive fields for different sublattices and lead
to avalanche-free reversal. This allows us to extract reversal
statistics for both sublattices during the reversal process.

\begin{figure}
  \begin{center}
    \includegraphics[width=\columnwidth]{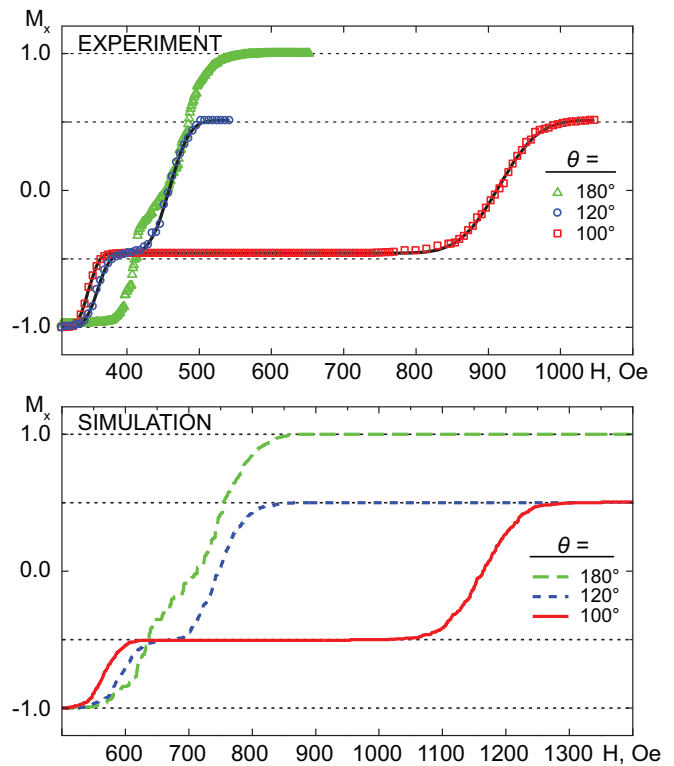}
    \caption{\label{fig3} Experimental (top) and theoretical (bottom)
      results of a uniaxial reversal along $100^{\circ}$\!\!\!,\,
      $120^{\circ}$\!\!\!,\, and $180^{\circ}$\!\!\!.\, Solid lines in
      the experimental results are fits of a superposition of error
      functions used to extract the reversal statistics for each
      sublattice.}
  \end{center}
\end{figure}

During reversal, the field is increased in steps of 5.0 Oe, and since
it can also effect the Lorentz imaging, the field must be reduced to 0
after each step, to capture an image at remanence.  Images are then
run through a MATLAB script that automatically determines the magnetic
configuration of each element; one image will give us one point on the
$\M(H)$ plot, as shown in the top panel of Fig.~\ref{fig3}.  These
results can then be compared with numerical simulations of an array of
interacting elements as described previously \cite{mellado2010}.
Simulated curves $\M(H)$ for a Gaussian distribution of coercive
fields with $\sigma/\overline{{H}_{c}} = 5\%$ are shown in the bottom
panel of Fig.~\ref{fig3}.  Details of the simulations will be
described elsewhere \cite{shen}.

To determine the disorder in our crystals directly, we fit the $\M(H)$
curves to the expected cumulative distribution function, a
superposition of error functions in the case when coercive fields have
Gaussian statistics:
\begin{equation}\label{fit}
M_{x} (H) = \sum_i
m_i\erf\left(\tfrac{H-\overline{{H}^{i}_{c}}}{\sigma_i\sqrt{2}}\right),
\end{equation}
where $M_{x}$ is the magnetization along the initial polarization
direction, $i$ indexes the sublattices in order of reversal, $m_{i}$
is the amount of magnetic moment carried by the sublattice,
$\overline{{H}^{i}_{c}}$ is the average coercive field, and
$\sigma_{i}$ is the standard deviation.  From fitting the reversal at
$\theta=100^{\circ}$\!\!\!,\, we obtain a disorder parameter,
$\sigma_{1} / \overline{{H}^{1}_{c}} = 3.3\%$, from the values
$\overline{{H}^{1}_{c}}=347 \ \mathrm{Oe}$ and $\sigma_{1}=11.3 \
\mathrm{Oe}$.  The second plateau at $\theta=100^{\circ}$ gives a
slightly higher disorder parameter of 4.7\%, from
$\overline{{H}^{2}_{c}}=915 \ \mathrm{Oe}$ and $\sigma_{2}=43.3 \
\mathrm{Oe}$.  We believe this is due to a rotational disorder from
element-to-element in the lattice, as expressed by the $\alpha$ offset
parameter from Equation \eqref{right}.  Disorder in $\alpha$ should
not be expected to contribute to $\sigma_{i}$ when
$\lvert\cos\left(\theta_{i}+\alpha\right)\rvert \approx 1$, which
occurs for $\sigma_{1}$ near $\theta=100^{\circ}$ in our system.  When
$\theta_{i} \approx \pm90^{\circ}$, the coercivity diverges and even
small rotational disorder would be expected to cause substantial
variations. From the reversal at $\theta=120^{\circ}$ degrees, we
obtain $\sigma_{1}/\overline{{H}^{1}_{c}}=3.7\%$, and
$\sigma_{2}/\overline{{H}^{2}_{c}}=4.6\%$, confirming the trend of
minimum disorder near the minimum $H^{i}_c$, and increasing
monotonically with $\theta$ away from that point.  We note that the
reversals at $\theta=180^{\circ}$ have a different character.  They
involve magnetic avalanches in which long chains of links reverse
together. We plan to address this question in a separate publication.

From this, we find that the static disorder is very low in our
samples, 3.3\%, which is a factor of 4-5 smaller than values for other
systems \cite{ladak2010,mengotti2011}.  This low amount of disorder is
a result of the geometry of our connected elements.  Here, charged
domain walls always exist at the vertices, even at remanence.  For a
given magnetic configuration, there is very little change in the
location and angle of the charged wall throughout the sub-lattice.
The same is not true for a disconnected lattice.  The process of
nucleating and injecting a domain wall is more sensitive to quenched
disorder in disconnected lattices, as the $\pm$2 domain walls must be
nucleated from the end of one of the elements instead of simply
injected, and any significant edge roughness may influence the
nucleation process.  This leads to an increase in the width of the
static disorder for a disconnected lattice.

We summarize the effect of the connected lattice on the disorder in
Table \ref{tab:disorder}, where we compare $\sigma/\overline{{H}_{c}}$
with two other experimental results, using different materials
\cite{ladak2010} and disconnected geometry \cite{mengotti2011}.  The
effect of the connected geometry is to decrease the disorder from 13\%
to 3.3\%. We acknowledge that the connected lattice presented herein
and the disconnected lattice used for comparison have been fabricated
by different research groups, but we have also performed simulations
that support the relative difference.  These micro-magnetic
calculations were performed using the OOMMF software package from NIST
\cite{oommf}.  Simulations were carried out for $180^{\circ}$
reversals on Y-junctions with random edge roughness, comparable to the
fabricated structures, and the results from an ensemble of
calculations are also shown in Table \ref{tab:disorder}, exhibiting a
similar relative relationship.

\begin{table}
  \begin{ruledtabular}
    \begin{tabular}{ccl}
      & \multicolumn{1}{c}{$\sigma_{1}/H_{c}^{1}$} & \multicolumn{1}{c} %
      {Composition \& Structure} \\
      \hline
      \multicolumn{1}{c}{\multirow{3}{*}{Experiment}} &
      \multicolumn{1}{c}{0.167} & Co, connected \cite{ladak2010} \\
      \multicolumn{1}{c}{} &
      \multicolumn{1}{c}{0.130} & NiFe, disconnected \cite{mengotti2011} \\
      \multicolumn{1}{c}{} &
      \multicolumn{1}{c}{0.033} & NiFe, connected, this work \\
      \hline
      \multicolumn{1}{c}{OOMMF} & %\multirow{2}{*}{OOMMF Simulations}} &
      \multicolumn{1}{c}{0.040} & NiFe, connected \\
      \multicolumn{1}{c}{Simulations} &
      \multicolumn{1}{c}{0.089} & NiFe, disconnected \\
    \end{tabular}
    \end{ruledtabular}
    \caption{\label{tab:disorder}Experimental reversal results and their
      associated disorder.  Theoretical results are shown for
      comparison. Theoretical results are averaged over 10 different
      simulations for each lattice.}
\end{table}
In addition to the lattice geometry, the material used in the
fabrication of the crystal is important.  Our crystals and most other
artificial spin ices have been made using Ni$_{80}$Fe$_{20}$, which
has a low crystal anisotropy, giving polycrystal films with uniform
magnetic properties.  This is not the same for cobalt, as used in
\cite{ladak2010}, which has a strong hexagonal anisotropy.  Magnetic
elements fabricated from polycrystal Co films would be expected to
have a broad distribution of coercivities.

By using a connected lattice of Ni$_{80}$Fe$_{20}$ elements, we have
alleviated both of these problems and created crystals with low
disorder.  Our crystal geometry is ideal in that it allows for a more
accurate representation of spin ice in frustrated magnetic materials.
For example, unlike other systems \cite{ladak2010,mengotti2011}, the
low-disorder system described herein rigorously obeys its
corresponding set of ice rules, a desirable feature given the
interaction energy of approximately $10^{4}$ K \cite{wang2006}.  This
presents the system as a more ideal model of frustration, providing
the hope of addressing outstanding problems, such as the occurrence of
a possible ground state in kagome ice \cite{moeller2009, chern}.

\begin{acknowledgments}
  This work was supported by the National Science Foundation under
  Grant Nos.  DMR-1056974 and DMR-0520491 and by NIST-DOC under
  Agreement 70NANB7H6177.
\end{acknowledgments}

\bibliographystyle{apsrev}
\bibliography{disorder}

\end{document}